\documentstyle[preprint,aps]{revtex}
\begin{document}

%
%

\title{ On the Liaison Between Superconductivity and
Phase Separation}

\author{S. Haas, E. Dagotto, A. Nazarenko}

\address{Department of Physics,
National High Magnetic Field Lab, and
Supercomputer Computations Research Institute,
Florida State University, Tallahassee, FL 32306}

\author{J. Riera}

\address{Physics Division, Oak Ridge National Laboratory,
Oak Ridge, Tennessee 37831}

\date{\today}
\maketitle

\begin{abstract}

Models of strongly correlated electrons that tend to phase separate
are studied
including a long-range 1/r repulsive interaction.
It is observed that charge-density-wave states
become stable as the strength of the
1/r term, ${\rm V_{coul}}$, is increased. Due to this effect,
the domain of stability of the superconducting phases that
appear near phase separation at ${\rm V_{coul} = 0}$ is not enlarged
by a 1/r interaction as naively expected.
Nevertheless, superconductivity exists
in a wide region of parameter
space,  even if phase separation is suppressed.
Our results have implications for some theories of the cuprates.

\end{abstract}


%
%

The presence of  charge modulations and phase separation in the
high-Tc compounds, as well as in other related materials,
has recently attracted considerable attention.
Experimental evidence has shown that the 38K superconductor
${\rm La_2 Cu O_{4+y}}$ (LCO)
has a regime where phase separation occurs,\cite{jor} which is adjacent to a
superconducting phase. Hole-rich and hole-poor phases coexist
at temperatures below ${\rm \sim 230K}$.
While a similar macroscopic phase separation does not occur in
${\rm La_{2-x} Sr_x Cu O_4}$ (LSCO), a microscopic segregation of
doped holes forming walls of hole-rich material separated by
undoped domains has been reported.\cite{cho}
Phase separation has also been observed in nonmetallic,
nonsuperconducting ${\rm La_{2} Ni O_{4+y}}$ (LNO),\cite{rice} while
electron diffraction techniques applied to ${\rm La_{2-x} Sr_{x} Ni O_{4+y}}$
have indicated the presence of charge modulations at low
temperatures.\cite{chen}
On the theoretical side, the description of electrons and holes in the
high-Tc cuprates is still under much discussion.
Among the many proposed theories, scenarios linking phase separation with
superconductivity have been discussed.\cite{emery,emery2,dago1,dago2}
In the widely studied two dimensional
t-J model phase separation exists at large J/t,
and in its vicinity
strong superconducting correlations
have been detected in the ground state.\cite{emery,dago1}
Actually, the presence of superconductivity close to phase
separation in electronic models seems a general phenomenon.\cite{dago2}
Fig.1(a,b,c) illustrates these ideas for the particular cases of the
one-dimensional (1D) and two-dimensional (2D) t-J
model, and also for the 2D t-U-V model i.e. the standard Hubbard model with a
nearest-neighbor density-density interaction. There is a growing
perception that phase separation and superconductivity are intimately
related in electronic models.\cite{emery2,dago2}

While a scenario linking phase separation and superconductivity is
appealing, the accumulation of charge in the hole-rich phase makes relevant
the long-range repulsive Coulomb
interaction among the carriers. Certainly
such interactions will render the phase separated regimes unstable in
Fig.1, while the superconducting phase, being uniform, should not be
strongly affected. Thus, a natural interesting possibility arises: can
it occur that once phase separation (PS) is suppressed by a long-range
force, the parameter regime formerly occupied by PS in Fig.1(a,b,c) is taken
over by
superconductivity? This effect would notably enhance the region of  stability
of the superconducting phase.
Actually, Kivelson and Emery predicted that high temperature
superconductivity is established once long-range Coulomb interactions
eliminate phase separation.\cite{emery2} However, preliminary studies in the
1D t-J model monitoring the pairing correlation functions showed that
superconductivity
was not enhanced by the 1/r term.\cite{troyer} It was there conjectured that
charge-density-wave (CDW) phases may be responsible for this effect.
More recently,
a study of a classical spin-one lattice gas model including
long-range repulsive interactions among spins observed that
the equivalent of ordered charge-modulated states become
stable.\cite{low}
Similar indications of CDW phases have also been reported in the context of
the three band Cu-O chain.\cite{riera}


The purpose of the present paper is to discuss a detailed analysis of
the competition between superconductivity, CDW and phase separation in
electronic models including
1/r repulsive forces. We considerably improve upon previous calculations
by using realistic models with mobile electrons not only in 1D but also
in 2D, and explicitly monitoring
the correlation functions corresponding to the three competing phases.
More specifically, the t-J model
have been studied using exact
diagonalization techniques, and the 2D t-U-V model, an effective model for
d-wave superconductors, has been also analyzed within
a mean-field approximation. We conclude that even in the presence of
mobile carriers, CDW phases indeed are stabilized by the
1/r term in a large region of parameter space, reducing the potential domain of
stability of the superconducting phase.
Nevertheless, we have observed that superconductivity is not strongly
suppressed
either (unless ${\rm V_{coul}}$ crosses some threshold), and thus
the fluctuating phase separation scenario for the cuprates,\cite{emery2} with
some
modifications to account for the  competition with CDW, remains a viable
possibility.
The strong competition between superconductivity, CDW, and phase
separation in electronic models
seems a general phenomenon.


The Hamiltonian for the ${\rm t-J}$ model with 1/r interactions has the form
$$
{\rm H_{tJ}=-t\sum_{<i,j>\sigma} (c^+_{i\sigma} c_{j\sigma}+h.c.)
  +J\sum_{<i,j>}({\bf S}_i\cdot{\bf S}_j-{1\over4} n_in_j)}$$
$$
{\rm + V_{coul}
\sum_{ij} {{n_i n_j}\over{ r_{ij} } }   },
\eqno(1)
$$
where the ${\rm c}$-operators are hole operators acting on non-doubly occupied
states,
${\rm r_{ij}}$ is the shortest distance between sites i and j,
and the rest of the notation is standard.
To analyze the ground state properties of this Hamiltonian
exact diagonalization techniques are used.\cite{review} Charge, spin and
pairing correlations are monitored
as a function of ${\rm J/t}$, ${\rm
V_{coul}/t}$, and $\langle n \rangle$. In 1D, the conformal field
theory parameter $k_{\rho}$ is also studied.\cite{ogata}.
This parameter is defined as
$ k_{\rho} = \pi v_c \langle n \rangle^2 \kappa /2$,
where the charge velocity
$v_c$ is obtained from the energy of the lowest
spin-singlet state in the subspace with momentum ${\rm k=2\pi/N}$ (${\rm N}$ is
the
number of sites), while the compressibility $\kappa$ at a given number of
particles ${\rm N_e}$  is obtained from the
ground state  energies with ${\rm N_e}$, ${\rm N_e +2}$ and ${\rm N_e - 2}$
electrons,
following standard procedures.\cite{ogata,troyer}
If $k_{\rho}
> 1$ the singlet pairing correlations decay the slowest against
distance, and thus are dominant in the ground state.
To discuss the results in both the 1D and 2D t-J model, for simplicity we will
use the
quarter filling density, $\langle n \rangle = 1/2$,
where the signals of superconductivity are clear even in 2D small clusters.
We have checked that for other densities (where superconductivity exists
near phase separation at ${\rm V_{coul} = 0}$), the results are
qualitatively similar as at $\langle n \rangle = 1/2$
(but with CDW states having a different modulation).

To start the analysis, it is
helpful to consider the atomic limit ${\rm V_{coul},J \gg t}$, where
intuition can be gained about the states that will compete with
superconductivity and
phase separation. This limit can be studied accurately on large clusters
and as anticipated from the spin one model,\cite{low}
several CDW phases were observed in the ground state as a function of ${\rm
J/V_{coul}}$ after the 1/r interaction makes the phase separated regime
unstable.
These CDW phases have
an increasing number of electrons in each microscopic cluster as
${\rm J/V_{coul}}$ grows, since J favors the formation of large spin
structures to gain antiferromagnetic energy (see Fig.2). Monitoring the
density-density correlation functions, we observed that the CDW states
are stable even for a finite hopping ${\rm t}$, and their
rough domain of stability is shown in Fig.2. Phase I is a
standard Wigner crystal. Phase II is a Wigner crystal of
pairs i.e. a regular distribution of charge 2e spin-singlets, similar to those
observed in the t-J-V model.\cite{emery,dago1} This state is stable
since the pairs take advantage of the short range effective attractive force
produced by J. Phase III has
clusters with four electrons, and as J increases the size of these
microscopic clusters also increases smoothly producing a cascade
of CDW phases. In the limit where J is the only
relevant scale, phase separation is recovered.\cite{comm77}

When the hopping amplitude t is nonzero, intermediate values
of J, and a small coupling ${\rm V_{coul}}$, a regime of superconductivity
exists in the
phase diagram as shown in Fig.2. Its boundaries were obtained
using exact diagonalization techniques on chains of 12 and 16
sites.\cite{review} This study shows that superconductivity dominates
(i.e. $k_{\rho} >1$) on a finite region of parameter space, while in the
bulk limit phase
separation exists only along the ${\rm V_{coul}=0}$ line in Fig.2.
In the pure t-J model, the compressibility diverges at the
boundary of phase separation, and thus $k_{\rho}$
on a finite cluster changes sign as shown in Fig.3a.
On the other hand, at
finite ${\rm V_{coul}}$, $k_{\rho}$ is a smooth function of J/t,
it becomes larger than one
on a small region, and then it smoothly converges to zero at
large J/t. To gain more intuition about the physical behavior of the
system, we also studied pairing correlations observing that in the
regime where
$k_{\rho}>1$, these correlations are indeed very strong in the ground
state and they continue having a large value,
even up to  ${\rm V_{coul}/t \sim 0.5}$ (with ${\rm J/t \sim 3}$) i.e. well
beyond the apparent stability regime signaled by $k_{\rho}$. This
curious effect shows that short distance
superconducting fluctuations may be relevant in a
wide region of parameter space, even if their asymptotic power-law
decay is not the dominant one.
On Fig.3b, the superconducting (pairing) and CDW (charge)
correlations (phase IV of Fig.2) summed over all distances
are plotted at a fixed ${\rm V_{coul} = 0.1t}$ against the exchange coupling.
The competition between the two phases is clear
showing explicitly
that even when phase separation is made unstable, the region left empty is
not necessarily taken over by superconductivity but by a crystal-like charge
modulation. Thus, our results confirm
the potential importance of charge-density-wave
phases in electronic models for the
cuprates.\cite{low,troyer,riera}

The analysis of the 2D t-J model is more complicated since there is
no analog of $k_{\rho}$, and the linear cluster sizes are
smaller than in 1D. Nevertheless, 2D is the dimension of interest for
the cuprates, and it deserves to be explored. We have carried out
an analysis of this case using $4\times 4$ clusters,
and monitoring the pairing and charge
correlations to set up the phase diagram.
In the atomic limit, which can be explored on larger clusters, the set of
stable CDW phases
is qualitatively different from the 1D case.
As observed before in the context of a classical spin-one
system\cite{low} and for the t-J model close to half-filling\cite{prelo},
striped phases (holes ordered in one dimensional chains
along the x or y axis) are
dominant in most of parameter space. As J increases, the number of
contiguous chains of electrons in the striped phases smoothly
increases. Away from the atomic limit,
a  numerical analysis of the pairing correlations similar to that
carried out for the pure 2D t-J model (see Ref.\cite{dago1}
for technical details)
suggests that superconductivity
is robust in a region of parameter space analogous in shape to that observed
in 1D. While the actual boundaries between the various phases in Fig.3c should
be
considered only as qualitative, nevertheless the trends are clear
from the numerics, and similar to those in 1D. Thus, our
study suggests that although phase separation is destabilized by the 1/r
correlation, the region left empty is not only replaced by the
neighboring superconductivity but also by CDW phases.


To show that this effect goes beyond the t-J model, and it may actually
affect all electronic models with coexisting superconductivity and phase
separation at ${\rm V_{coul}=0}$, here we consider the t-U-V model.
Its Hamiltonian is
$$
{\rm H =-t\sum_{<i,j>\sigma} (c^+_{i\sigma} c_{j\sigma}+h.c.)
  +  U  \sum_{i} n_{i\uparrow} n_{i\downarrow}
  +  V \sum_{<i,j>} n_{i} n_{j}  }$$
$$
{\rm + V_{coul} \sum_{ij} {{n_i n_j}\over{ r_{ij} } }   },
\eqno(2)
$$
where the notation is standard (for  details see
Ref.\cite{dago2}). The phase diagram at ${\rm V_{coul}=0}$ and half-filling was
shown in
Fig.1c using a mean-field approximation. For simplicity, let us
consider ${\rm U/t=0}$ which is at the center of the stability region of
the d-wave superconducting phase, and an attractive V term which
can be effectively produced by spin-polaronic correlations
in a more realistic Hamiltonian (results at other values of U/t are
qualitatively similar).
As was shown before,\cite{dago2} this model at half-filling has a stable d-wave
superconducting phase with order parameter
${\rm \Delta_{q}=\frac{\Delta_{0}}{2}(\cos{q_{x}}-\cos{q_{y}}) }$, where
$\Delta_0$ is obtained through the self-consistent equation
${\rm \sum_{\bf k \rm} \frac{V_{eff}(\bf p\rm -\bf k)\rm \Delta_{\bf k}}{
\sqrt{\epsilon_{\bf k}^{2}+
\Delta^{2}_{\bf k}} } = \Delta_{\bf p}}$,
with ${\rm V_{eff}(\bf q \rm ) = \frac{1}{2N}\sum_{\bf r\rm}
exp(i\bf qr \rm ) [ V(\delta_{\bf r \rm ,\hat{\bf x}}+\delta_{\bf r \rm ,-
\hat{\bf x}}+\delta_{\bf r \rm ,\hat{\bf y}}+\delta_{\bf r
\rm ,-\hat{\bf y}}) }$ ${\rm +\frac{V_{coul}}{r} ] }$.
The ground state energy of this condensate is
${\rm E_{GS} =
\frac{-{\Delta_0}^2}{4(V+V_{coul})}-\frac{1}{N}\sum_{\bf k\rm}
\sqrt{\epsilon_{k}^{2}+
\Delta^{2}_{k}}+\frac{E_{PS}}{2N} },$
where ${\rm E_{PS}}$ is the energy of the phase separated state (half the
lattice empty,
the other half doubly occupied).\cite{dago2}
The energies of the CDW and phase separated states can also be written
in closed  form, but it is impractical to quote
them here.
We have also introduced a cutoff of four lattice spacings for
technical reasons dealing with the 1/r interactions. This makes phase
separation stable on a finite region for any finite cutoff.
A straightforward comparison of the energies of the competing
states was enough to establish the mean-field phase diagram shown in Fig.4a. As
it
happened for the t-J model, in this case also the d-wave phase is
eventually suppressed by CDW states. Note that there is a large region of
stability for the d-wave phase but this may be a consequence of the
approximations made in evaluating the energy of the different phases.
However, we do not expect
qualitative changes in the phase diagram given improved approximations.
Note that the effect of CDW and phase separation is usually neglected in
studies
of models of d-wave superconductors, which Fig.4b suggests to be a
dangerous practice.


In this paper, we have studied the competition between superconductivity,
phase separation and charge ordered states in models of correlated
electrons both in 1D and 2D.
Increasing the strength of the 1/r repulsive
interaction, we found that superconductivity does not take over the
entire region left empty by phase separation, since CDW
states become competitive in energy as observed in the
spin-one lattice gas model.\cite{low} Nevertheless the
superconducting phase survives in the presence of the Coulomb
interaction as long as its strength is not too large.
In Fig.4b, we summarize our results with
a schematic representation of the influence of
long-range interactions over an idealized system which at ${\rm
V_{coul}=0}$ has both superconductivity and phase separation. ``g''is the
strength of a generic coupling constant which produces an attractive
short-range force in the system. It is
interesting to attempt to locate some of the known cuprates in
different places of this phase diagram (Fig.4b). LCO seems to have a small
effective Coulomb interaction or screening length since phase
separation exists in this compound, while LSCO may have
intermediate values of these parameters that suppress phase separation. LNO
(due
to its small mobility caused by the large Ni spin) seems located in the CDW
region of Fig.4b.
Summarizing, in electronic models where effective short-range attractive forces
are
operative, superconductivity competes for stability with phase
separation if the repulsive 1/r interaction is neglected, and with CDW states
if 1/r
interactions are included.

\medskip
E. D. is supported by the Office of Naval Research under
grant ONR N00014-93-0495.  J. R. was supported by
the Department of Energy (DOE) Office of Scientific Computing, by the DOE under
contract No. DE-AC05-84OR21400, and under contract No.
DE-FG05-87ER40376 with Vanderbilt University.

\medskip

\vfil\eject

%
%

{\bf Figure Captions}

\begin{enumerate}

\item
Illustration of the presence of phase separation near superconductivity
in electronic models $without$ 1/r interactions:
(a) Phase diagram of the 1D t-J model as a function of
the electronic density $\langle n \rangle$ and J/t
from Ref.\cite{ogata}. PS denotes
``phase separation'', $k_{\rho}>1$
($k_{\rho}<1$) implies that singlet superconducting (spin and charge)
correlations decay the slowest in the ground state; (b)
Schematic phase diagram of the 2D t-J model taken
Ref.\cite{dago2}. The notation is similar to (a) with $x = 1 - \langle n
\rangle$, ``SC'' denoting a
mostly d-wave superconducting region, ``PM'' a paramagnetic region, and
``AF'' antiferromagnetism;
(c) Mean-field phase diagram
of the 2D t-U-V model at half-filling (Ref.\cite{dago2}).
${\rm SS}$, ${\rm DS}$, ${\rm CDW}$, ${\rm
SDW}$, and ${\rm PS}$ denote s-wave
superconductivity, d-wave superconductivity,
charge-density-wave, spin-density-wave, and phase separation, respectively.

\item
Phase diagram of the 1D t-J model with long-range 1/r interactions
(see Eq.(1)) at density $\langle n \rangle = 1/2$ obtained from the
analysis of a chain with 16 sites.
The region $k_{\rho} > 1$ is where superconductivity
dominates in the ground state.
When $k_{\rho} < 1$ and at small
${\rm V_{coul}}$, SDW and CDW competes .
At large values of ${\rm V_{coul}}$, or to the right
of the superconducting region, the ground state is dominated by CDW
order (states I, II, III, ... having different charge modulations).
The open (full) circles denote holes (electrons).
The size of each microscopic cluster in the CDW state increases
as J/t  increases, and the many CDW phases after IV are not shown.
The dashed lines are tentative boundaries separating the different
charge orders. Phase separation (PS) is only stable at ${\rm V_{coul}
=0}$ in the bulk limit.

\item
(a) $k_{\rho}$ in the 1D  t-J model, at density $\langle n
\rangle = 1/2$ on a 16 sites
chain. The solid line corresponds to ${\rm V_{coul}=0.00}$ and
shows the divergence of $k_{\rho}$ at the point of phase separation. On
the other hand at ${\rm V_{coul}=0.15t}$, $k_{\rho}$ remains finite;
(b) Behavior of the pairing and charge susceptibilities ($\chi_{\rm sup}$
and $\chi_{\rm CDW}$, respectively) vs J/t at fixed ${\rm
V_{coul}=0.10t}$; (c) Phase
diagram of the 2D t-J model at density $\langle n \rangle = 1/2$
obtained using a 16 sites square cluster. SC
is (d-wave) superconductivity, PM a paramagnetic region, PS phase
separation,
I is the standard Wigner crystal, II a Wigner crystal of pairs, and
III and IV are shown in the figure. Note the strong similarities with Fig.2.

\item
(a) Mean-field phase diagram of the 2D t-U-V model including 1/r interactions
at half-filling.
${\rm PS}$ denotes phase separation (stable over a
finite region since in this calculation since the 1/r was truncated at four
lattice spacings), ${\rm d-wave}$ is the superconducting phase, while ${\rm
FL}$ denotes a Fermi-liquid region obtained by turning off the
superconducting
order parameter in the mean-field calculation. I is a CDW Wigner
crystal, II
is a one column  striped phase (i.e. phase III of Fig.3c), and III is
shown; (b) Generic representation of the phase diagram
of electronic models with competing short-range attractive and
long-range repulsive forces. For details see the text.

\end{enumerate}


\begin{references}
\bibitem{comm77} Strictly speaking, in 1D there are no ordered states. However,
the CDW states have very large correlation lengths and
for all practical purposes they behave as having LRO on finite lattices.
\end{references}
\end{document}